\documentclass[twocolumn, prd, aps, showpacs, preprintnumbers]{revtex4-2}

\usepackage{tikz}
\usepackage{feynmp-auto}
\usepackage{amsmath}
\usepackage{todonotes}

\usepackage[utf8]{inputenc}
\usepackage[T1]{fontenc} 
\usepackage{array}
\usepackage{bbm}
\usepackage{subcaption}
\usepackage{soul}	
\usepackage{xcolor}
\usepackage{graphicx}
\usepackage{epsfig}
\usepackage{color}
\usepackage{slashed}
\usepackage{comment}
\usepackage{epstopdf}
\usepackage{xspace}
\usepackage{mathrsfs}
\usepackage[euler]{textgreek}
\usepackage{amsmath}
\usepackage{amsthm}
\usepackage{amsfonts}
\usepackage{amssymb}
\usepackage{graphicx}
\usepackage[font={small}]{caption}
\usepackage{subcaption}
\usepackage{float}
\usepackage{multirow}
\usepackage[hang, flushmargin,bottom]{footmisc} 
\usepackage{placeins}
\usepackage{enumitem}
\usepackage{slashed}
\usepackage{bbm}
\usepackage{appendix}
\usepackage{tabularx}
\usepackage{siunitx}

\usepackage{titlesec}
\titleformat{\chapter}[display]
  {\normalfont\LARGE\bfseries}
  {\chaptertitlename\ \thechapter}{5pt}{\LARGE}
  \titlespacing*{\chapter}{0pt}{-20pt}{35pt}
\usepackage{bigstrut}
\usepackage{pst-node}
\usepackage{pstricks}
\usepackage{physics}
\usepackage{mathtools}
\usepackage{setspace}
\usepackage{fancyhdr}
\usepackage[makeroom]{cancel}
\usepackage{feyn}
\newcommand{\be}{\begin{equation}}
\newcommand{\ee}{\end{equation}}
\newcommand{\bes}{\begin{equation*}}
\newcommand{\ees}{\end{equation*}}

\usepackage{hyperref}% http://ctan.org/pkg/hyperref
\hypersetup{%
  colorlinks = true,
  linkcolor  = blue,
  citecolor  = blue,
  urlcolor   = blue,
}
\usepackage{soul}
\usepackage{xpatch}
\makeatletter
\xpretocmd{\todo}{\@bsphack}{}{}
\xapptocmd{\todo}{\@esphack}{}{}
\makeatother

\newcommand{\beq}{\begin{equation}}
\newcommand{\eeq}{\end{equation}}

\usepackage{tikz}

\definecolor{green}{HTML}{008000}
\definecolor{goldenrod}{HTML}{DAA520}
\definecolor{magenta}{HTML}{FF00FF}
\definecolor{silver}{HTML}{C0C0C0}
\definecolor{indigo}{HTML}{4B0082}
\definecolor{skyblue}{HTML}{87CEEB}
\definecolor{darkgoldenrod}{HTML}{B8860B}
\definecolor{orange}{HTML}{FFA500}
\definecolor{yellow}{HTML}{FFFF00}
\definecolor{saddlebrown}{HTML}{8B4513}
\definecolor{blue}{HTML}{0000FF}
\definecolor{turquoise}{HTML}{40E0D0}
\definecolor{yellow}{HTML}{FFFF00}
\definecolor{white}{HTML}{FFFFFF}
\definecolor{whitesmoke}{HTML}{F5F5F5}
\definecolor{hotpink}{HTML}{FF69B4}

\newcommand{\myComment}[1]{}

\label{sec:axion}     % Section 2: The QCD Axion
\label{sec:DFSZ}      % Section 3: DFSZ Mechanisms
\label{sec:KSVZ}      % Section 4: KSVZ Mechanisms
\label{sec:neutrinos} % Section 5: Neutrinos from the QCD Axion
\label{sec:summary}   % Section 6: Summary

\begin{document}
\title{\Large{The QCD Axion and Neutrino Masses}}
\author{Hridoy Debnath, Pavel Fileviez P{\'e}rez, Grace Miller}
%\email{fileviez@physics.wisc.edu}
\affiliation{
Physics Department and Center for Education and Research in Cosmology and Astrophysics, Case Western Reserve University, Cleveland, OH 44106, USA}
\email{hxd253@case.edu, pxf112@case.edu, gmm139@case.edu}
\date{\today}

\begin{abstract}
The strong CP problem and the origin of neutrino masses are among the clearest motivations for physics beyond the Standard Model. The Peccei--Quinn mechanism offers an elegant dynamical solution to the strong CP problem, predicting the QCD axion, while neutrino masses point to new degrees of freedom or interactions beyond the Standard Model. We explore the possibility that these two phenomena arise from a common origin by identifying the Peccei--Quinn symmetry-breaking scale with the scale responsible for neutrino mass generation.
We perform a systematic and unified analysis of this connection in both DFSZ and KSVZ realizations of the Peccei--Quinn mechanism, considering the Type-I, Type-II, and Type-III seesaw mechanisms, the Zee radiative model, and the colored seesaw. For each scenario, we determine the axion couplings to photons and neutrinos and examine their phenomenological consequences. A robust prediction of all the frameworks considered is the existence of axion-neutrino couplings.
We show that these interactions can significantly affect the axion decays. If the QCD axion makes up the observed dark matter, these couplings also lead to a large monochromatic neutrino flux from axion decays, opening a complementary path to test these theories in future neutrino experiments. 
\end{abstract}

\maketitle
%
%%%%%%%%%%%%%%%%%%%%%%%%
\section{INTRODUCTION}
%%%%%%%%%%%%%%%%%%%%%%%%
The Standard Model (SM) of particle physics stands as one of the most
successful theoretical frameworks ever constructed, describing the
fundamental constituents of matter and their interactions with
extraordinary precision. Yet, despite its remarkable achievements,
it leaves two of the most profound questions in
contemporary particle physics and cosmology unanswered: the strong CP-problem and the
origin of neutrino masses. The fact that these two problems both
require new physics beyond the SM, and that both point
toward new symmetries and new energy scales, raises the compelling
possibility that they may share a common theoretical origin. This
work is devoted to a systematic and unified exploration of that
possibility.

The strong CP problem arises from the structure of Quantum
Chromodynamics (QCD) itself. The SM permits a CP-violating
interaction in the strong sector whose coefficient, the parameter
$\bar\theta$, receives contributions from both the QCD vacuum angle
and the phases of the quark mass matrices. Experimental measurements
of the neutron electric dipole moment constrain this parameter to be
extraordinarily small, $\bar\theta \lesssim 10^{-10}$
\cite{Abel:2020pzs,Pospelov:1999mv}, a level of fine-tuning that
has no explanation within the SM. One of the solutions to this puzzle
was proposed by Peccei and Quinn (PQ) \cite{Peccei:1977hh,Peccei:1977ur},
who introduced a new global anomalous symmetry, $U(1)_{\rm PQ}$,
that is spontaneously broken at a high energy scale $f_a$. The
associated pseudo-Nambu--Goldstone boson, the QCD axion~\cite{Weinberg:1977ma,Wilczek:1977pj}, acquires a potential from
non-perturbative QCD effects whose minimum dynamically relaxes
$\bar\theta$ to zero, solving the strong CP problem without any
fine-tuning. Beyond its role in solving the strong CP problem, the
QCD axion is one of the most compelling dark matter candidates
\cite{Preskill:1982cy,Abbott:1982af,Dine:1982ah}, with a mass
inversely proportional to the PQ symmetry-breaking scale
and a rich phenomenology accessible to a broad range of laboratory,
astrophysical, and cosmological searches.

The origin of neutrino masses constitutes a more  fundamental
challenge. The observation of neutrino flavor oscillations by the
Super-Kamiokande \cite{Fukuda:1998mi} and SNO \cite{Ahmad:2001an}
collaborations established beyond doubt that neutrinos are massive,
in direct contradiction with the original formulation of the SM. The absolute scale of neutrino masses is further
constrained by cosmological observations \cite{Aghanim:2018eyx} and
direct kinematic measurements \cite{Aker:2021gma}, with the sum of
neutrino masses bounded to be below the electron-volt scale.
Explaining such tiny masses requires the introduction of new degrees
of freedom or new interactions beyond the SM. 

The most
natural and widely studied frameworks for neutrino mass generation
are the seesaw mechanisms, in which the smallness of neutrino masses
is attributed to a new lepton number violating
scale. In Type-I seesaw
\cite{Minkowski:1977sc,Mohapatra:1979ia,GellMann:1980vs,
Yanagida:1979as,Schechter:1980gr}, heavy right-handed
neutrinos with a large mass suppress light
neutrino masses through the seesaw relation. The Type-II seesaw
\cite{Konetschny:1977bn,Lazarides:1980nt,Mohapatra:1980yp,Schechter:1981bd}
achieves the same result through the vacuum expectation value of a
scalar electroweak triplet, while the Type-III seesaw
\cite{Foot:1988aq} employs fermionic electroweak triplets. An
alternative and equally well-motivated possibility is the radiative
generation of neutrino masses, exemplified by the Zee model
\cite{Zee:1980ai}, in which Majorana masses arise at the one-loop
level. A particularly
elegant colored variant of the radiative mechanism was proposed in Ref.~\cite{FileviezPerez:2009gr}, in which
color-octet fermions and scalars generate neutrino masses at one-loop level.

The possibility that the PQ symmetry and the symmetry
responsible for neutrino mass generation share a common origin is
theoretically well-motivated and has been explored in a number of
contexts
\cite{Shin:1987xj,Ballesteros:2016xej,Dias:2014osa,
Ballesteros:2016euj,Salvio:2015cja,Clarke:2015bea,
Bertolini:2015ola,Ahn:2015pia,Arason:1990pu,deGiorgi:2023tvn,Batra:2023erw,Babu:2026yqp}.
The central observation is that the same scalar singlet responsible
for breaking the PQ symmetry can also generate the heavy
mass scale entering the neutrino sector, naturally relating the
axion decay constant to the seesaw scale. This identification has
profound phenomenological consequences: it constrains the axion
parameter space from above by the requirement that the
PQ scale is related to the canonical seesaw scale.
In these scenarios one
predicts characteristic axion couplings to neutrinos that are
proportional to the neutrino masses and suppressed by the
PQ scale. Despite the growing
literature on specific realizations of this connection, a
comprehensive and unified comparison of the different neutrino mass
mechanisms within both the 
Dine--Fischler--Srednicki--Zhitnitsky~(DFSZ)
\cite{Dine:1981rt,Zhitnitsky:1980tq} and the
Kim--Shifman--Vainshtein--Zakharov~(KSVZ)
\cite{Kim:1979if,Shifman:1979if} frameworks has been
lacking.

In this work, we provide a systematic and unified
study of the connection between the QCD axion and the origin of
neutrino masses. We analyze this connection within both the DFSZ and the KSVZ realizations of the PQ
mechanism, considering five representative neutrino mass generation
scenarios: the Type-I, Type-II, and Type-III seesaw mechanisms, the
Zee radiative model, and the colored seesaw. For each
framework, we compute the electromagnetic and color
anomaly coefficients that determine the axion--photon coupling and
derive the axion coupling to neutrinos. 

Within the DFSZ framework,
we show that the Type-I seesaw, Type-II seesaw, and Zee mechanism
leave the anomaly coefficients unchanged with respect to the minimal
DFSZ model, since the additional fields responsible for neutrino
masses are either gauge singlets or scalars and do not contribute to
the triangle anomaly. In contrast, the Type-III seesaw introduces
fermionic electroweak triplets that carry electromagnetic charge and
modify the anomaly ratio $E/N$, leading to a distinctive and
measurable shift in the axion--photon coupling. 
We show for the first time the predictions in all possible scenarios with colored seesaw and show that in these scenarios using only the fields needed to generate neutrino masses at one-loop level, one can implement the Peccei-Quinn mechanism for the QCD axion. Notice that in other scenarios one can add extra fields to achieve the main goal but the answer is not unique as in the case of the colored seesaw scenarios.

We discuss two
further classes of observables that are generic predictions of
frameworks in which the PQ scale is identified with the
seesaw scale. First, the axion has coupling to
neutrinos proportional to the neutrino mass divided by the
PQ scale, which determines the axion decay channels,
lifetime, and branching fractions between the diphoton and dineutrino
modes. Second, assuming that the QCD axion constitutes the observed
dark matter, these axion--neutrino interactions predict a
large monochromatic neutrino flux from axion dark matter decays that lies
well above both the Cosmic Neutrino Background and the solar neutrino
flux over a wide range of energies, providing a complementary probe
of the QCD axion in future neutrino experiments. 

This article is organized as follows. In
Section~\ref{sec:axion} we review the QCD axion and its couplings.
In Section~\ref{sec:DFSZ} we discuss the DFSZ realizations of the
PQ mechanism combined with the Type-I, Type-II, Type-III,
and Zee neutrino mass mechanisms, deriving the corresponding anomaly
coefficients and axion--photon couplings. In Section~\ref{sec:KSVZ}
we analyze the KSVZ framework and the colored seesaw construction. In
Section~\ref{sec5} we discuss the axion--neutrino coupling,
the axion lifetime and branching fractions, and the predicted
monochromatic neutrino flux from axion dark matter decays. We summarize our main findings in Section~\ref{sec:summary}. 
%%%%%%%%%%%%%%%%%%%%%%%%%
\section{THE QCD AXION}
%%%%%%%%%%%%%%%%%%%%%%%%
\label{sec:axion}
The SM allows for a renormalizable CP-violating interaction
in QCD,
\begin{equation}
    \mathcal{L}_\theta = \theta \, \frac{g_s^2}{32\pi^2}
    G^a_{\mu\nu} \tilde{G}^{a\mu\nu},
    \label{eq:Ltheta}
\end{equation}
where $G^a_{\mu\nu}$ is the gluon field-strength tensor and
$
    \tilde{G}^{a\mu\nu} = \frac{1}{2}
    \epsilon^{\mu\nu\alpha\beta} G^a_{\alpha\beta}.
    \label{eq:Gtilde}
$
After taking into account the phases appearing in the quark mass
matrices, the physical parameter controlling strong CP violation is given by
\begin{equation}
    \bar{\theta} = \theta + \arg\det(M_u) + \arg\det( M_d).
    \label{eq:thetabar}
\end{equation}
As is well-known, the experimental limit on the neutron electric dipole moment
requires, $\bar{\theta} \lesssim 10^{-10}$, which constitutes the
strong CP problem. Notice that the quark mass matrices are related to the observed CP-violation in the quark sector given by the Cabibbo-Kobayaski-Maskawa matrix, $V_{CKM}$.
The PQ mechanism provides an elegant
dynamical solution by introducing a global anomalous $U(1)_{\rm PQ}$
symmetry that is spontaneously broken at an energy scale $f_a$.
The associated pseudo-Nambu--Goldstone boson is the QCD axion.
Because the PQ symmetry is anomalous under QCD, the axion couples
to gluons according to
\begin{equation}
    \mathcal{L} \supset \frac{\alpha_s}{8\pi}
    \frac{a}{f_a} G^a_{\mu\nu} \tilde{G}^{a\mu\nu}.
    \label{eq:aGG}
\end{equation}
Non-perturbative QCD effects generate an effective potential for the
axion whose minimum dynamically cancels the effective CP-violating
parameter $\bar\theta$, thereby solving the strong CP problem. The axion mass is
inversely proportional to the PQ symmetry-breaking scale~\cite{Gorghetto:2018ocs},
\begin{equation}
    m_a \simeq 5.7 \; \mu{\rm eV}
    \left( \frac{10^{12} \; {\rm GeV}}{f_a} \right).
    \label{eq:axionmass}
\end{equation}
At energies below the electroweak
scale, the effective axion interactions can be written as
\begin{eqnarray}
    \mathcal{L} &\supset& \frac{\alpha_s}{8\pi} \frac{a}{f_a} G^a_{\mu\nu} \tilde{G}^{a\mu\nu}
    + \frac{1}{4} g_{a\gamma\gamma} \, a F_{\mu\nu} \tilde{F}^{\mu\nu}
    \nonumber \\
    &+& \sum_f \frac{C_f}{2f_a} \partial_\mu a \,
    \bar{f} \gamma^\mu \gamma^5 f,
    \label{eq:Leff}
\end{eqnarray}
where the couplings to photons and fermions depend on the ultraviolet
realization of the PQ symmetry. The axion--photon coupling is
conventionally parameterized as
\begin{equation}
    g_{a\gamma\gamma} = \frac{\alpha_{\rm em}}{2\pi f_a}
    \left( \frac{E}{N} - 1.92 \right),
    \label{eq:gagg}
\end{equation}
where $E$ and $N$ denote the electromagnetic and color anomaly
coefficients, respectively. The ratio $E/N$ therefore distinguishes
different invisible axion models.
Two broad classes of invisible axion models have been extensively
studied. In the DFSZ construction, the SM Higgs sector
is enlarged by an additional Higgs doublet together with a complex
scalar singlet carrying PQ charge, allowing the SM
fermions to carry PQ charges. In contrast, the KSVZ mechanism
introduces heavy vector-like colored fermions that generate the
anomalous axion--gluon coupling after being integrated out, while
the SM fermions remain neutral under the PQ symmetry.

\begin{figure}[t]
    \centering
    \includegraphics[width=0.99\linewidth]{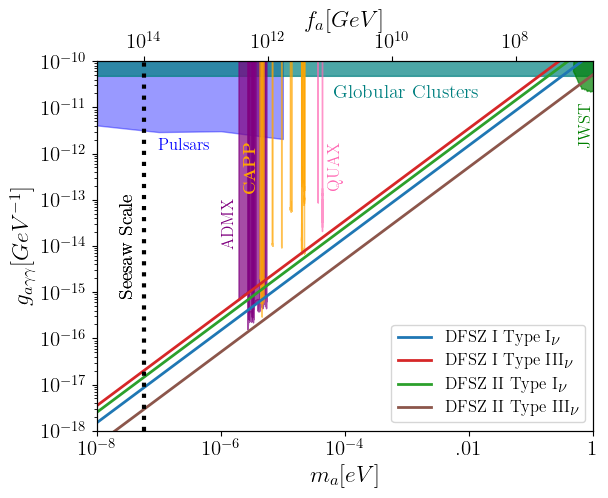}
    \caption{Axion photon coupling as a function of axion mass for the DFSZ scenarios.  The teal shaded region indicates the bounds from globular clusters \cite{Ayala:2014pea, Dolan:2022kul}, the green region from the JWST experiment \cite{gemmell2026reexaminingsensitivityjwstdecaying},  the purple region for ADMX \cite{asztalos2010squid, ADMX:2018gho, ADMX:2019uok, Bartram_2021, Goodman_2025, Carosi_2025}, grey for QUAX \cite{Di_Vora_2023,Rettaroli_2024}, orange for CAPP \cite{Lee:2020cfj, Jeong_2020, Yoon_2022, Kim_2023, Yi_2023,Yang_2023, kim2024experimentalsearchinvisibledark, Bae:2024kmy, CAPP:2024dtx, Adair_2022}, and blue for pulsars \cite{Noordhuis:2022ljw}. The black dotted line provides a constraint from neutrino masses when we assume that the PQ symmetry breaking cannot be larger than the canonical seesaw scale. Notice that the cavity searches as ADMX, CAPP, and QUAX, and the bounds from JWST assume that the QCD axion is responsible for $100\%$ of the dark matter relic density. Here we use the bounds listed in Ref.~\cite{AxionLimits}.}
    \label{fig:DFSZ_photon}
\end{figure}
In this work, we are interested in models where the PQ symmetry is
directly connected to the origin of neutrino masses. Such
constructions naturally relate the axion decay constant to the scale
responsible for lepton number violation or the seesaw scale and
predict characteristic axion couplings to neutrinos. In the
following sections we discuss these possibilities within both the
DFSZ and KSVZ frameworks. 
For a review about the QCD axion see Ref.~\cite{DiLuzio:2020wdo}.
For the connection between the QCD axion and neutrino masses, see~\cite{Shin:1987xj,Ballesteros:2016euj,Dias:2014osa,Ballesteros:2016xej,Salvio:2015cja,Clarke:2015bea,Bertolini:2014aia,Ahn:2015pia,PhysRevD.43.2337,Ma:2017vdv}. Notice that most of these studies are in the context of Type I and Type II seesaw mechanics.
%%%%%%%%%%%%%%%%%%%%%%%%%%%%
\section{DFSZ MECHANISMS}
%%%%%%%%%%%%%%%%%%%%%%%%%%
\label{sec:DFSZ}
The DFSZ framework provides one of the most economical realizations of the PQ mechanism. Unlike KSVZ models, where the QCD anomaly is generated by heavy vector-like colored fermions, the DFSZ construction extends the scalar sector of the SM while allowing the ordinary quarks and leptons to carry PQ charges. Consequently, the axion couples directly to SM fermions at tree level, and its interactions are determined by the PQ charge assignments. The minimal DFSZ model contains two Higgs doublets,
\begin{equation}
H_u \sim (1,2,-1/2), \qquad
H_d \sim (1,2,1/2),
\end{equation}
together with a complex scalar singlet
$
S \sim (1,1,0),
$
whose vacuum expectation value spontaneously breaks the global
$U(1)_{\rm PQ}$ symmetry,
$
\langle S \rangle = v_S/\sqrt{2},
$
giving rise to the QCD axion as the associated pseudo-Nambu--Goldstone boson.
The Yukawa interactions responsible for quark masses, together with the scalar
interaction that communicates PQ symmetry breaking to the Higgs sector, are given by
\begin{eqnarray}
\mathcal{L}_{\rm DFSZ}
&\supset&
Y_u \,\overline{q}_L H_u u_R
+
Y_d \,\overline{q}_L H_d d_R \nonumber \\
&+&
\lambda \,
H_u^T i\sigma_2 H_d\, {S^\dagger}^2
+\mathrm{h.c.}.
\end{eqnarray}
The charged lepton sector is more model dependent. Depending on the PQ charge
assignments, the charged leptons may couple either to $H_d$ or to $H_u$,
leading to the familiar DFSZ-I and DFSZ-II realizations~\cite{DiLuzio:2020wdo}. Although both
implement the PQ solution to the strong CP problem, they predict
different electromagnetic anomaly coefficients and therefore different axion--photon couplings.
One of the most attractive features of the DFSZ framework is that the same
scalar singlet responsible for breaking the PQ symmetry can also generate the
masses of the heavy fields required for neutrino mass generation. In this case,
the PQ symmetry-breaking scale and the seesaw scale become naturally related,
\begin{equation}
M_{\rm seesaw} \sim v_S \sim f_a,
\end{equation}
establishing a direct connection between the solution of the strong CP problem
and the origin of neutrino masses. 

Depending on the particle content, the PQ
anomaly coefficients may either remain unchanged or receive additional
contributions from new fermionic degrees of freedom.
\begin{table}[t]
\centering
%\begin{subtable}{0.45\textwidth}
%\centering
%\caption{First table}
\begin{tabular}{|c|c|}
\hline
\textbf{Model } & \textbf{E/N}\\ \hline
\makebox[11em][c]{DFSZ I - Type I, II or Zee} & 8/3 \\ \hline
\makebox[8em][c]{DFSZ I - Type III} & 11/3 \\ \hline
\makebox[8em]{DFSZ II - Type I, II or Zee} & 2/3 \\ \hline
\makebox[8em]{DFSZ II - Type III} & 5/3 \\ \hline
\end{tabular}
%\end{subtable}
\caption{Values of $E/N$ for the DFSZ scenarios.}
\label{tableI}
\end{table}
Overall, the DFSZ framework provides a particularly elegant realization of the
interplay between the PQ mechanism and neutrino mass generation.
While the Type-I, Type-II, and Zee models preserve the standard DFSZ
predictions for the axion couplings, the Type-III realization modifies the
electromagnetic anomaly coefficient through the presence of additional
electroweak fermions. The corresponding values of the anomaly coefficients,
$E/N$, for the different scenarios are summarized in
Table~\ref{tableI}. Now, we can consider several scenarios:
\begin{itemize}
\item Type-I seesaw:
The simplest realization introduces at least two right-handed neutrinos~\cite{Minkowski:1977sc,Mohapatra:1979ia,GellMann:1980vs,
Yanagida:1979as,Schechter:1980gr},
\begin{equation}
\nu_R \sim (1,1,0),
\end{equation}
whose Majorana masses originate from the vacuum expectation value of the PQ
scalar,
\begin{eqnarray}
\mathcal{L}_{\rm DFSZ}^{\rm I}
&\supset&
Y_\nu\,\overline{\ell}_L H_u \nu_R
+
\lambda_R \nu_R^T C \nu_R S
+\mathrm{h.c.}. 
\end{eqnarray}

After spontaneous symmetry breaking, the singlet scalar generates the heavy
Majorana mass matrix,
$
M_R=\lambda_R v_S \sqrt{2},
$
and the usual Type-I seesaw mechanism gives
$
m_\nu \simeq -m_D M_R^{-1} m_D^T.
$
Since the right-handed neutrinos are singlets under the SM gauge
group, they do not contribute to either the color or electromagnetic PQ
anomalies. Consequently, the anomaly coefficients $N$ and $E$, and therefore
the axion couplings to gluons and photons, remain identical to those of the
minimal DFSZ model. The new phenomenological feature is the derivative coupling
between the axion and neutrinos. This scenario has been studied in Refs.~\cite{Shin:1987xj,Ballesteros:2016euj,Dias:2014osa,Salvio:2015cja}.
\item Type-II seesaw:
In the Type-II realization the SM is extended by a scalar triplet~\cite{Konetschny:1977bn,Magg:1980ut,Lazarides:1980nt,Mohapatra:1980yp,Schechter:1981bd},
$
\Delta \sim (1,3,1),
$
and the relevant interactions are
\begin{eqnarray}
\mathcal{L}_{\rm DFSZ}^{\rm II}
&\supset&
y_\nu\,
\ell_L^T C i\sigma_2 \Delta \ell_L \nonumber \\
&+&
\lambda_\Delta
H_d^T i\sigma_2
\Delta^\dagger
H_d S 
+\mathrm{h.c.}. 
\end{eqnarray}
After PQ symmetry breaking, the scalar singlet induces the small vacuum
expectation value of the triplet, leading to Majorana neutrino masses. Since the additional particle responsible for neutrino masses is a scalar,
there is no modification of either the QCD or electromagnetic anomaly
coefficients. Therefore, the axion-photon coupling remains identical to that of
the corresponding minimal DFSZ-I or DFSZ-II model.
This scenario was studied in Ref.~\cite{Bertolini:2015ola}.
\item Type-III seesaw:
The Type-III realization replaces the right-handed neutrinos by fermionic
electroweak triplets~\cite{Foot:1988aq},
$
\rho_R \sim (1,3,0),
$
whose masses are generated by the PQ-breaking scalar,
\begin{equation}
\mathcal{L}_{\rm DFSZ}^{\rm III}
\supset
Y_\nu\,
\overline{\ell}_L H_u \rho_R
+
\lambda_\rho
\rho_R^T C \rho_R S
+\mathrm{h.c.}.
\end{equation}
In contrast to the previous scenarios, however, the fermionic
triplets carry electroweak quantum numbers and contribute to the
electromagnetic anomaly. Consequently, the ratio $E/N$ differs from that of the
minimal DFSZ construction, leading to a modified axion-photon coupling while
leaving the color anomaly unchanged. This provides a distinctive experimental
signature capable of distinguishing the Type-III realization from the Type-I
and Type-II scenarios. 
This scenario was briefly mentioned in Ref.~\cite{Ahn:2015pia}.
\item Zee mechanism:
In the Zee model an additional charged scalar~\cite{Zee:1980ai},
$
\delta^+ \sim (1,1,1),
$
together with an extra Higgs doublet one generates Majorana neutrino
masses at the one-loop level. Since the Zee mechanism introduces only scalar degrees of freedom beyond the
SM fermion content, the PQ anomaly coefficients remain identical to
those of the minimal DFSZ model. Consequently, the predictions for the axion
couplings to gluons and photons are unchanged, although the theory still
predicts axion--neutrino interactions once the PQ symmetry is directly linked
to the origin of neutrino masses.
It is important to mention that to have a realistic Zee mechanism for neutrino masses one needs two Higgs doublets coupled to charged leptons. This scenario was investigated in Ref.~\cite{Arason:1990pu}.
\end{itemize}
Fig.~\ref{fig:DFSZ_photon} summarizes the predictions for the axion--photon coupling, $g_{a\gamma\gamma}$, as a function of the axion mass for the different DFSZ realizations in which the PQ symmetry is directly connected to the origin of neutrino masses. As expected, the Type-I, Type-II, and Zee scenarios reproduce the standard DFSZ predictions because the additional fields responsible for neutrino mass generation do not modify the electromagnetic anomaly coefficient, $E/N$. In contrast, the Type-III seesaw introduces electroweak fermion triplets that contribute to the electromagnetic anomaly, leading to a different value of $E/N$ and consequently shifting the prediction for $g_{a\gamma\gamma}$ while leaving the QCD anomaly coefficient unchanged. 

The figure~\ref{fig:DFSZ_photon} also compares these theoretical predictions with current astrophysical and laboratory constraints, including the bounds from Pulsars, globular cluster, the recent limits derived from \textit{JWST} observations, and the exclusion regions from ADMX and other haloscope experiments. Furthermore, the black dotted line indicates the theoretical upper bound obtained by requiring that the PQ symmetry-breaking scale does not exceed the canonical seesaw scale. The distinct prediction bands associated with the different values of the anomaly ratio $E/N$ demonstrate that a future precision measurement of the axion--photon coupling could distinguish between the various neutrino mass generation mechanisms, thereby providing a powerful probe of the underlying realization of the PQ symmetry. Notice that mainly the Global Clusters, JWST and ADMX experiments rule out a fraction of the parameter space. In particular, the global cluster bound helps us to to find the allowed lower bound on the PQ scale in most of the scenarios. The JWST bound is more relevant for DFSZ II - Type III$_\nu$ model.
%%%%%%%%%%%%%%%%%%%%%%%%%%%%%%
\section{KSVZ MECHANISMS}
%%%%%%%%%%%%%%%%%%%%%%%%%%%%%%
\label{sec:KSVZ}
The KSVZ mechanism provides an alternative
realization of the Peccei--Quinn mechanism. In contrast to DFSZ models, the
SM fermions are neutral under the global $U(1)_{\rm PQ}$ symmetry.
Instead, the QCD anomaly responsible for solving the strong CP problem is
generated by heavy vector-like colored fermions. After the spontaneous breaking
of the PQ symmetry, these heavy fermions induce the effective axion-gluon
interaction through the chiral anomaly.
The minimal KSVZ construction introduces a complex scalar singlet,
$
S\sim(1,1,0),
$
whose vacuum expectation value 
breaks the PQ symmetry spontaneously. The model also contains a
vector-like pair of colored fermions,
$
\Psi_L$, and $\Psi_R,
$
which transform identically under the SM gauge symmetry but carry
different PQ charges. In these models the relevant interactions are given by
\begin{equation}
\mathcal{L}_{\rm KSVZ}
\supset
y_\Psi \overline{\Psi}_L\Psi_R S
+\mathrm{h.c.},
\end{equation}
leading to the vector-like masses,
$
M_\Psi=y_\Psi v_S/\sqrt2.
$
Since the SM fermions do not carry PQ charge, the axion has no
tree-level couplings to ordinary quarks and leptons. Instead, its interactions
with gluons, photons, and matter are generated radiatively after integrating
out the heavy colored fermions. Consequently, the phenomenology of KSVZ models
is largely determined by the gauge quantum numbers of the heavy vector-like
fermions.
Throughout this work we consider three simple standard realizations,
\begin{itemize}
\item  {\rm KSVZ~I:}
$D_{L,R}\sim(3,1,-1/3)$, 
\item {\rm KSVZ~II:}
$U_{L,R}\sim(3,1,2/3)$, 
\item {\rm KSVZ~III:}
$Q_{L,R}\sim(3,2,1/6)$,
\end{itemize}
whose anomaly coefficients are summarized in Table~\ref{tableII}. For simplicity, we assume Type I seesaw for these models. Notice that if we study Type III seesaw, the new triplets will change the coupling to photons as we discussed in the previous section. 
Although these models
provide identical solutions to the strong CP problem, they predict different
electromagnetic anomaly coefficients and therefore different axion-photon
couplings. 

A particularly attractive possibility is to identify the heavy colored fermions
required by the KSVZ mechanism with the particles responsible for generating
neutrino masses. In this way, the PQ scale simultaneously determines
the mass of the heavy colored states and the scale of lepton number violation.
Among the known realizations, the colored seesaw mechanism provides one of the
most elegant examples of this connection.
That is the main reason we discuss the predictions in colored seesaw in great detail.
\begin{table}[t]
\centering
\begin{tabular}{|c|c|}
\hline
\textbf{Model } & \textbf{E/N}\\ \hline
Colored Seesaw  I & 0 \\ \hline
Colored Seesaw  II & 16/9 \\ \hline
KSVZ I & 2/3 \\ \hline
KSVZ II & 8/3 \\ \hline
KSVZ III & 5/3 \\ \hline
\end{tabular}
\caption{Values of $E/N$ for the different KSVZ scenarios.}
\label{tableII}
\end{table}
%%%%%%%%%%%%%%%%%%%%%%%%%%%%%%%%%%%%%%%%%%%%%
\subsection{Colored Seesaw and the QCD Axion}
%%%%%%%%%%%%%%%%%%%%%%%%%%%%%%%%%%%%%%%%%%%%%%%
One can consider a mechanism for neutrino masses at one-loop level without assuming an extra symmetry and avoid any stable colored or electrically charged fields.
This is the idea behind the colored seesaw mechanism~\cite{FileviezPerez:2009gr}.
In this context, one generates Majorana neutrino masses at the quantum level by
introducing a scalar color octet,
\begin{equation}
\Phi\sim(8,2,1/2),
\end{equation}
and fermionic color octets,
\begin{equation}
\rho_L \sim(8,R,0),
\end{equation}
where $R=1$ or $3$ denotes the $SU(2)_L$ representation. The relevant
interactions are
\begin{eqnarray}
- \mathcal{L}
&\supset&
Y^\nu
\ell_L^T C i\sigma_2
\rho_L \Phi
+
\lambda_\rho
{\rm Tr}
(\rho^T_L C\rho_L)S \nonumber \\
&+&
\frac{\lambda}{2}
{\rm Tr}
(\Phi^\dagger H)^2
+\mathrm{h.c.}.
\end{eqnarray}
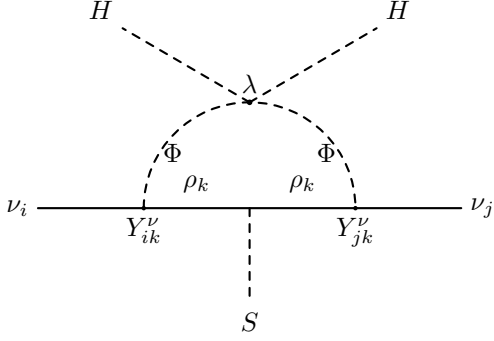
\begin{figure}[t]
        \centering
\begin{tikzpicture}[scale=0.7, line cap=round, line join=round]

  % Main horizontal line
  \draw[thick] (-4,0) -- (4,0);

  % End labels
  \node[left] at (-4,0) {$\nu_i$};
  \node[right] at (4,0) {$\nu_j$};

  % Key points on horizontal line
  \coordinate (L) at (-2,0);
  \coordinate (R) at (2,0);
  \coordinate (C) at (0,0);
  \coordinate (T) at (0,2);

  % Labels under the lower vertices
  \node[below] at (L) {$Y_{ik}^{\nu}$};
  \node[below] at (R) {$Y_{jk}^{\nu}$};

  % Top point lambda
  \fill (T) circle (1.5pt);
  \node[above] at (T) {$\lambda$};

  % Lower small dots
  \fill (L) circle (1.2pt);
  \fill (R) circle (1.2pt);

  % Dashed semicircle
  \draw[dashed, thick] (L) arc[start angle=180,end angle=0,radius=2];

  % rho_k labels inside the semicircle
  \node at (-1,0.45) {$\rho_k$};
  \node at (1,0.45) {$\rho_k$};

  % Phi labels near the sides
  \node at (-1.45,1.05) {$\Phi$};
  \node at (1.45,1.05) {$\Phi$};

  % Dashed lines from lambda to H points
  \draw[dashed, thick] (T) -- (-2.4,3.4);
  \draw[dashed, thick] (T) -- (2.4,3.4);

  % H labels
  \node[above left] at (-2.4,3.4) {$H$};
  \node[above right] at (2.4,3.4) {$H$};

  % Dashed vertical line downward to S
  \draw[dashed, thick] (C) -- (0,-1.8);
  \node[below] at (0,-1.8) {$S$};

\end{tikzpicture}
\caption{Colored seesaw mechanism.}
\label{fig:CS}
\end{figure}
Once the singlet scalar, $S$, develops a vacuum expectation value, the fermionic
octet acquires a Majorana mass,
$
M_\rho=
\lambda_\rho v_S \sqrt2,
$
while light neutrino masses are generated through the one-loop diagram in Fig.~\ref{fig:CS},
\begin{equation}
(M_\nu)_{ij}
\frac{\lambda v^2}{16\pi^2}
\sum_k
Y^\nu_{ik}
Y^\nu_{jk}
I(M_{\rho_k},M_\Phi),
\end{equation}
where $I(M_{\rho_k},M_\Phi)$ denotes the loop function~\cite{FileviezPerez:2009gr}.
The same scalar field responsible for generating the heavy octet mass also
breaks the PQ symmetry. Consequently, the axion decay constant and
the scale entering the colored seesaw are naturally related,
\begin{equation}
f_a=\frac{v_S}{N},
\end{equation}
where $N$ denotes the QCD anomaly coefficient.
%%%%%%%%%%%%%%%%%%%%%%
\begin{figure}[t]
        \centering
\includegraphics[width=0.99\linewidth]{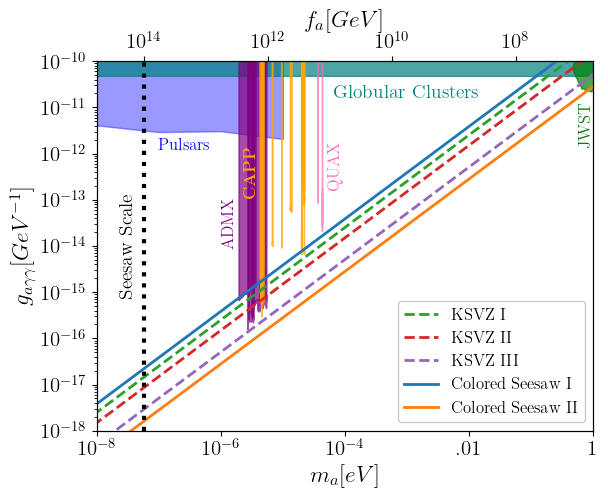}
        \caption{ Axion photon coupling as a function of axion mass for the various KSVZ scenarios. The teal shaded region indicates the bounds from globular clusters \cite{Ayala:2014pea, Dolan:2022kul}, the green region from the JWST experiment \cite{gemmell2026reexaminingsensitivityjwstdecaying},  the purple region for ADMX \cite{asztalos2010squid, ADMX:2018gho, ADMX:2019uok, Bartram_2021, Goodman_2025, Carosi_2025}, grey for QUAX \cite{Di_Vora_2023,Rettaroli_2024}, orange for CAPP \cite{Lee:2020cfj, Jeong_2020, Yoon_2022, Kim_2023, Yi_2023,Yang_2023, kim2024experimentalsearchinvisibledark, Bae:2024kmy, CAPP:2024dtx, Adair_2022}, and blue for pulsars \cite{Noordhuis:2022ljw}. The black dotted line provides a constraint from neutrino masses when we assume that the PQ symmetry breaking cannot be larger than the canonical seesaw scale.}
\label{aggg}
        \label{fig:KSVZ_photon}
    \end{figure}
In this framework two minimal realizations are particularly interesting:
\begin{itemize}
\item Colored Seesaw I:
$\rho_L \sim(8,1,0)$.
\item Colored Seesaw II:
$\rho_L \sim(8,3,0)$.
\end{itemize}
These two scenarios predict different values of the anomaly ratio $E/N$ and
therefore distinct axion-photon couplings, while both successfully explain the
origin of neutrino masses and solve the strong CP problem. The corresponding
predictions are summarized in Table~\ref{tableII}.
Here we consider only one copy of the new fermions. 
However, the ratio between $E$ and $N$ does not change if we add several copies of the fermionic field.
See Appendix~\ref{A} for the detailed calculation of the coefficients $N$ and $E$.
\begin{figure}[t]
    \centering
    \includegraphics[width=0.9\linewidth]{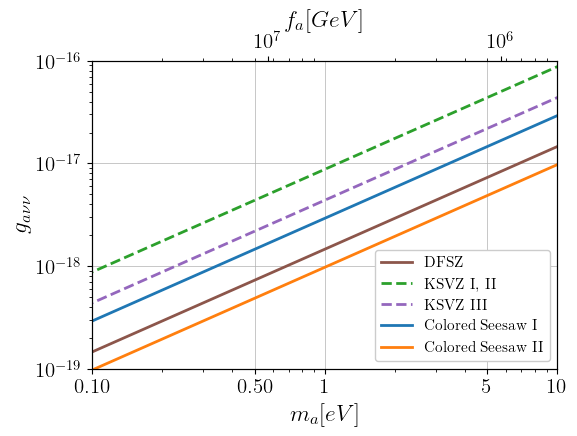}
    \caption{Axion neutrino coupling as a function of axion mass for the heaviest neutrino, $m \approx 0.05$ eV.}
    \label{fig:axion_neutrino}
\end{figure}
Fig.~\ref{fig:KSVZ_photon} shows the predictions for the axion--photon coupling, $g_{a\gamma\gamma}$, as a function of the axion mass for the different KSVZ realizations discussed in this work. In contrast to the DFSZ framework, the axion--photon coupling in KSVZ models is entirely determined by the gauge quantum numbers of the heavy vector-like colored fermions responsible for generating the QCD anomaly. Consequently, the different choices of fermion representations lead to distinct values of the anomaly ratio $E/N$, resulting in clearly separated predictions for $g_{a\gamma\gamma}$. In particular, the minimal KSVZ-I, KSVZ-II, and KSVZ-III models, as well as the two colored seesaw realizations, populate different regions of the parameter space, reflecting the different electromagnetic anomaly coefficients associated with each construction. 

The figure~\ref{fig:KSVZ_photon} displays the current astrophysical constraints from Pulsars, the globular clusters and recent JWST observations, together with the exclusion limits from ADMX and other haloscope experiments. As in the DFSZ case, the black dotted line represents the theoretical constraint obtained by requiring that the PQ symmetry breaking scale does not exceed the canonical seesaw scale, thereby emphasizing the connection between the origin of neutrino masses and the solution to the strong CP problem. The broad separation among the predicted axion--photon couplings illustrates that future laboratory searches with improved sensitivity will have the capability to discriminate between the different KSVZ realizations and, in particular, to test colored seesaw scenarios in which the PQ symmetry simultaneously explains the strong CP problem and generates neutrino masses. 
We would like to emphasize that the predictions in the colored seesaw scenarios are quite different, predicting the ratio:
\begin{equation}
g_{a\gamma\gamma}^{I}/g_{a\gamma\gamma}^{II} \sim 14.
\end{equation}
Therefore, the bounds are much stronger in colored seesaw I.
Notice the importance of the Global Clusters, JWST and ADMX bounds, which are able to rule out a fraction of the parameter space.
%
%%%%%%%%%%%%%%%%%%%%%%%%%%%%%%%%%%%%%%%%%%
\section{NEUTRINOS FROM THE QCD AXION}
\label{sec5}
%%%%%%%%%%%%%%%%%%%%%%%%%%%%%%%%%%%%%%%%%
A common feature of these models is that the QCD axion has a coupling to SM neutrinos. This coupling reads as
\begin{equation}
- {\cal{L}} \supset c_\nu \frac{\partial_\mu a }{ 4 v_S} \bar{\nu} \gamma^\mu \gamma^5 \nu.
\end{equation}
Here the coefficient $c_\nu$ is one for the KSVZ models, while for the DFSZ models it is function of $\tan \beta=v_u/v_d$. For example, in the simple case where one has Type I seesaw and DFSZ mechanisms, one finds that $c_\nu=1-4/(1 + \tan^2 \beta)$. However, the exact dependence is model dependent and here for simplicity we show the generic predictions when $c_\nu=1$. 

Fig.~\ref{fig:axion_neutrino} shows the predicted axion--neutrino coupling, $g_{a\nu \nu}$ for the heaviest neutrino, as a function of the axion mass. For illustration we assume $c_\nu=1$. A generic prediction of all scenarios discussed in this work is that the axion couples derivatively to neutrinos once the PQ symmetry is identified with the symmetry responsible for generating neutrino masses. After expressing the coupling in terms of the physical neutrino masses, one finds $g_{a\nu_i\nu_i}= c_{\nu} m_{\nu_i}/(N f_a)$, implying that the coupling increases linearly with the axion mass through the relation $m_a\propto f_a^{-1}$. We show the numerical predictions for all KSVZ and DFSZ models considered.
Although these interactions are highly suppressed by the large PQ scale, they constitute a robust and model-independent prediction of frameworks in which the PQ symmetry is directly linked to the origin of neutrino masses. These couplings determine the partial decay widths of sufficiently heavy axions into neutrino pairs and therefore play a central role in the predictions for the axion lifetime, branching fractions, and the monochromatic neutrino flux from axion dark matter discussed below. 
\begin{figure}[t]
    \centering
    \begin{subfigure}[b]{0.5\textwidth}
        \centering
        \includegraphics[width=\textwidth]{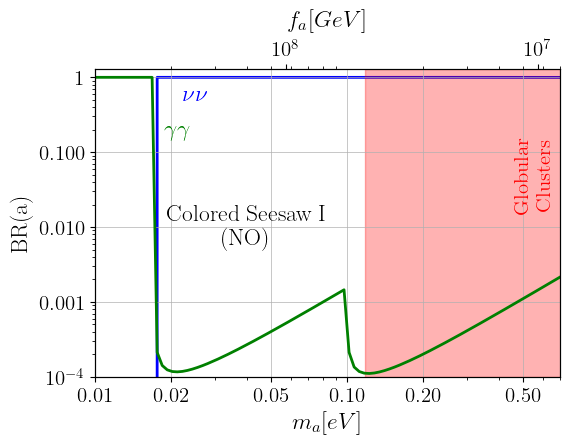}
        \caption{}
        \label{subfig:A1}
    \end{subfigure}  
%    \hfill
    \begin{subfigure}[b]{0.5\textwidth}
        \centering
        \includegraphics[width=\textwidth]{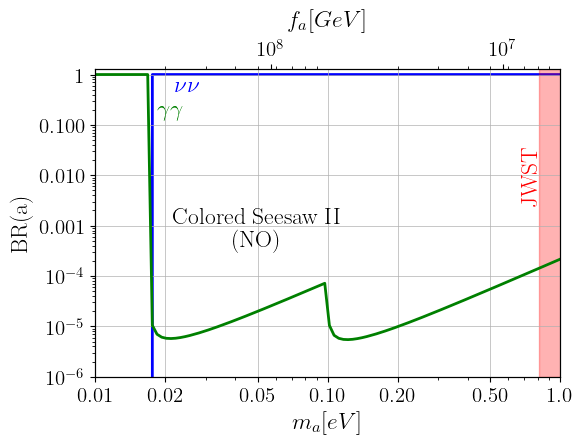}
        \caption{}
        \label{subfig:A3}
    \end{subfigure}   
    \caption{Branching ratios of axion decays as a function of axion mass. a) colored seesaw I with normal-order neutrino mass spectrum, $m_1=0,m_2 \approx 8.6\times 10^{-3} \text{eV}, m_3 \approx 0.05 \ \text{eV}$ b) colored seesaw II with normal-order neutrino mass spectrum. }
     \label{fig:BRaxion1}   
\end{figure}
\begin{figure}
%    \par\vspace{0.5cm} % Creates space before the next row
    \begin{subfigure}[b]{0.5\textwidth}
        \centering
        \includegraphics[width=\textwidth]{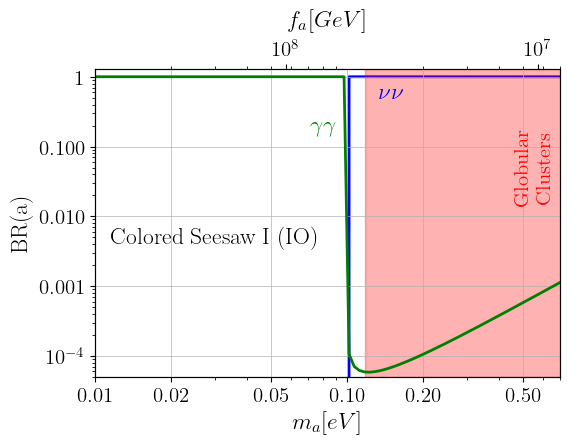}
        \caption{}
        \label{subfig:A3}
    \end{subfigure} 
    \begin{subfigure}[b]{0.5\textwidth}
        \centering       \includegraphics[width=\textwidth]{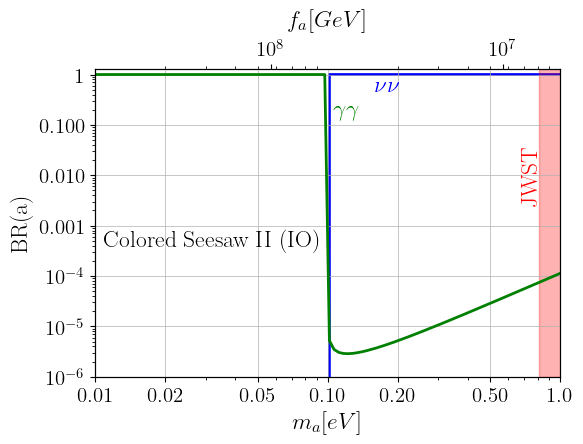}
        \caption{}
    \end{subfigure} 
    \caption{Branching ratios of axion decays as a function of axion mass. 
    c) colored seesaw I with inverted order neutrino mass spectrum, $m_3 =0$, $m_1 \approx 0.0492 $ eV and $m_2 \approx 0.05$ eV d) colored seesaw II with inverted order neutrino mass spectrum.}
 \label{fig:BRaxion2}   
 \end{figure}

\begin{figure}[t]
    \centering
    \begin{subfigure}[b]{0.45\textwidth}
        \centering
        \includegraphics[width=\textwidth]{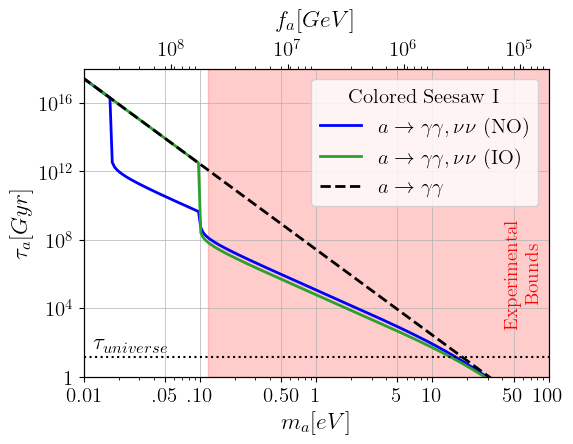}
        \caption{}
        \label{subfig:A1}
    \end{subfigure}  
%    \hfill
    \begin{subfigure}[b]{0.45\textwidth}
        \centering
        \includegraphics[width=\textwidth]{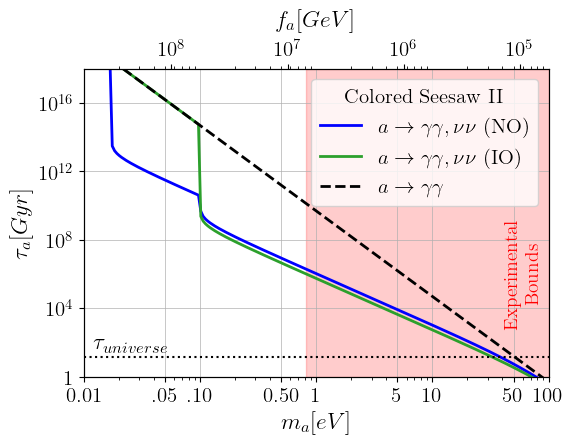}
        \caption{}
        \label{subfig:A3}
    \end{subfigure}          
%    \par\vspace{0.5cm} % Creates space before the next row
\caption{Axion lifetime as a function of axion mass assuming the normal-ordered neutrino mass spectrum. In a) we show the predictions in colored seesaw I, while in b) we have the predictions in colored seesaw II.}
\label{fig:lifetime}
 \end{figure}
The decay width for the channel with two photons is given by
\begin{equation}
    \Gamma(a \to \gamma\gamma) = \frac{\alpha_{em}^2}{256 \pi^3} (E/N -1.92)^2 \frac{m_a^3}{f_a^2},
\end{equation}
while for the decays into two neutrinos one has
\begin{equation}
    \Gamma(a \to \nu_i \nu_i) = \frac{m_a}{16 \pi}\  \left|\frac{c_{\nu} m_{\nu_i}}{N f_a}\right|^2\sqrt{1-\frac{4 m_{\nu_i}^2}{m_a^2}}.
\end{equation}
\begin{figure}[b]
    \centering
    \begin{subfigure}[b]{0.45\textwidth}
        \centering
        \includegraphics[width=\textwidth]{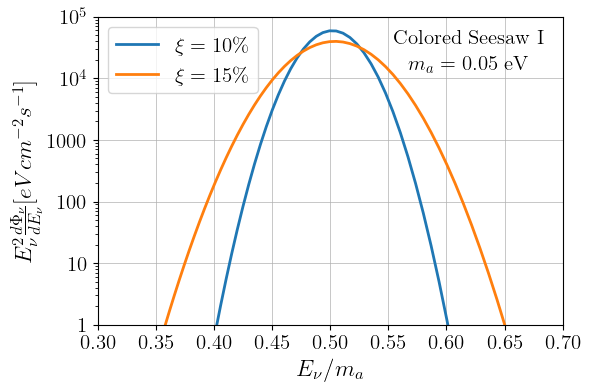}
        \caption{}
        \label{subfig:E1}
    \end{subfigure}  
%    \hfill
    \begin{subfigure}[b]{0.45\textwidth}
        \centering
        \includegraphics[width=\textwidth]{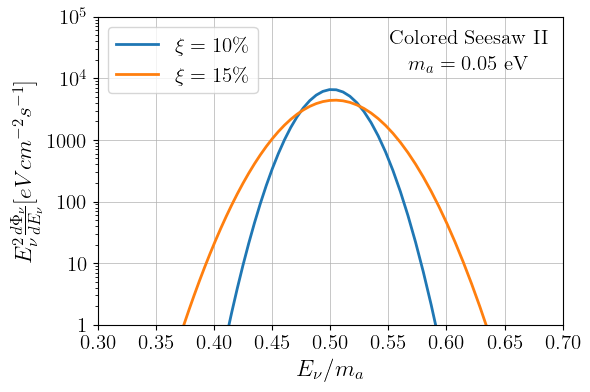}
        \caption{}
        \label{subfig:E3}
    \end{subfigure}          
%    \par\vspace{0.5cm} % Creates space before the next row
\caption{Differential neutrino energy spectrum from the $a\to \nu\nu$ decay as a function of $E_{\nu}/m_a$ for different energy resolutions. Here, we assume $m_a=0.05$ eV and  normal ordered mass spectrum for neutrinos. In a) we show the predictions in colored seesaw I, while in b) we have the predictions for colored seesaw II.}
\label{fig:energySpectrum}
 \end{figure}
Figs.~\ref{fig:BRaxion1} and \ref{fig:BRaxion2} shows the branching fractions of the QCD axion as a function of its mass for the two colored seesaw realizations and for both normal-ordering (NO) and inverted-ordering (IO) neutrino mass spectra. Since the axion--neutrino coupling is proportional to the neutrino mass, the relative importance of the different decay channels is sensitive to the neutrino mass hierarchy. In the low-mass region, the decay into two photons is the dominant channel because decays into neutrinos are kinematically forbidden. Once the threshold for the production of neutrino pairs is crossed, the branching fractions into the heavier neutrino mass eigenstates increase rapidly and eventually dominate the total decay width. In the normal-ordering scenario, the largest contribution originates from the decay into the heaviest neutrino state, $\nu_3\nu_3$, while the decay into $\nu_1\nu_1$ is absent due to the assumption of a massless lightest neutrino. Conversely, in the inverted-ordering case, the dominant channels are $a\rightarrow\nu_1\nu_1$ and $a\rightarrow\nu_2\nu_2$, whereas the decay into $\nu_3\nu_3$ is absent. Although the branching fractions are largely controlled by the neutrino mass spectrum, small quantitative differences between the Colored Seesaw~I and Colored Seesaw~II realizations arise from their distinct anomaly coefficients, which determine the axion--photon coupling and therefore modify the relative importance of the diphoton decay mode. Consequently, measurements of the decay pattern of a sufficiently heavy QCD axion could provide complementary information on both the neutrino mass ordering and the underlying realization of the PQ symmetry.

Fig.~\ref{fig:lifetime} displays the predicted axion lifetime as a function of the axion mass for the two colored seesaw realizations, assuming a normal-ordering neutrino mass spectrum. The lifetime is determined by the competition between the decay channels into two photons and neutrino pairs. For axion masses below the neutrino-pair production threshold, the decay $a\rightarrow\gamma\gamma$ dominates, resulting in extremely long lifetimes that greatly exceed the age of the Universe over the entire QCD axion mass range of interest. Once the decays into neutrinos become kinematically allowed, the total decay width increases due to the additional contributions from the axion--neutrino interactions, leading to a corresponding reduction in the lifetime. Since the axion--photon coupling depends on the electromagnetic anomaly coefficient, the Colored Seesaw~I and Colored Seesaw~II scenarios exhibit slightly different lifetimes, with the differences becoming most apparent in the region where the photon and neutrino decay channels are comparable. Nevertheless, over most of the parameter space the lifetime remains extraordinarily large, confirming that the QCD axion is a possible dark matter candidate even in the presence of axion--neutrino interactions. 

The expected differential neutrino flux at Earth, for each neutrino flavor, resulting from the decay of  axion dark matter with mass $m_a$ and lifetime $\tau_a$ is given by~\cite{Arguelles:2022nbl}
\begin{equation}
    \frac{d \Phi_\nu}{dE}= \frac{1}{4 \pi} \frac{1}{m_a \tau_a} \frac{1}{3} \frac{dN}{dE} D(\Omega),
\end{equation}
where  
\begin{equation}
    \frac{dN}{dE}=2 \delta(E-m_a/2).
\end{equation}
The D factor depends on the DM distribution of the galaxy. The D-factor is obtained by integrating the dark matter density distribution, $\rho(x)$, along the line of sight and over the observed solid angle $\Delta \Omega$: 
\begin{equation}
    D = \int d\Omega \int \rho(x)\  dx.
\label{D}
\end{equation}
The value of the D factor for the all sky is $2.65 \times 10^{23}$ GeV $\text{cm}^{-2}$ sr~\cite{Arguelles:2022nbl}.

 In Fig.\ref{fig:energySpectrum} we show the differential neutrino energy spectrum from the axion dark matter decay for colored seesaw I and II scenarios. The predicted spectra are shown for detector energy resolutions of $10\%$ and $15\%$ for a benchmark scenario of axion mass 0.05 eV. Here we take into account the detector's finite energy resolution using a Gaussian function, and the spectrum function $dN/{dE}$ can be written as 
\begin{equation}
    \frac{dN}{dE} =2 \int_0^\infty dE_0 \  \delta (E_0 -m_a/2) \ G(E_\nu,\xi/\omega,E_0),
\end{equation}
with
\begin{equation}
    G(E_\nu,\xi/\omega,E_0) = \frac{1}{\sqrt{2\pi}E_0 (\xi/\omega)} \ e^{-\frac{(E_\nu -E_0)^2}{2 E_0^2(\xi/\omega)^2}},
\end{equation}
where $\xi$ defines the energy resolution of the detector and $\omega= 2\sqrt{2\text{log}2}$ determines the full width at half maximum with the standard deviation and the standard deviation is given by $\sigma_0 = E_0 \xi/\omega$. 
 The comparison between the two scenarios shows that the predicted neutrino spectrum changes significantly for different values of the anomaly coefficients of E and N. 
 Notice that the flux is large and
 these spectral features can provide a distinctive signature for probing axion dark matter at future neutrino observations.

Fig.~\ref{fig:neutrino_flux} displays the predicted monochromatic neutrino flux from the decay of axion dark matter as a function of the neutrino energy $E$, for all seesaw realizations discussed above. The figure also shows, for comparison, the flux of the Cosmic Neutrino Background~(CNB), indicated by the black line, and the well-established solar
thermal neutrino flux~\cite{Vitagliano:2019yzm}, shown in dark blue. The predictions
for the colored seesaw~I, and
colored seesaw~II, are shown in blue
and orange, respectively. 
The overall normalization of the flux is inversely proportional to the axion
lifetime, $\tau_a$, and directly proportional to the local dark matter density
integrated along the line of sight through the D-factor, as given in
Eq.~(\ref{D}). All seesaw scenarios predict fluxes that are
clearly separated from both the CNB and the solar neutrino background. 
\begin{widetext}
\begin{center}
\begin{figure}[h]
    \centering
    \includegraphics[width=0.8\linewidth]{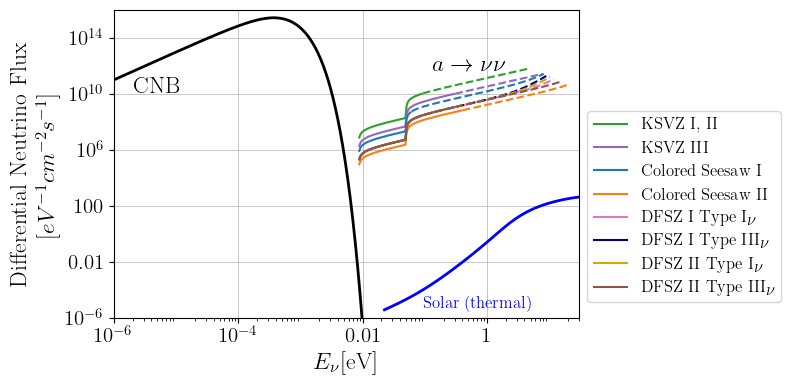}
    \caption{Differential neutrino flux from axion dark matter as a function of the neutrino energy, $E_\nu = m_a/2$. The differential flux for the cosmic neutrino background (CNB) is shown by the black line, the  predictions for the solar (thermal) neutrinos is shown by the blue line and taken from ~\cite{Vitagliano:2019yzm}. The differential neutrino flux from the axion decays are shown by the colored lines, as labeled in the legend. The dashed portions indicate the JWST and globular cluster restrictions.}
    \label{fig:neutrino_flux}
\end{figure}
\end{center}
\end{widetext}

The direct detection of the Cosmic Neutrino Background (CNB), one of the last unverified predictions of the Big Bang theory, has become a realistic experimental objective thanks to the proposed PTOLEMY experiment~\cite{PTOLEMY:2019hkd,ptolemy2018}. PTOLEMY aims to detect relic neutrinos through the neutrino capture process on tritium, $\nu_e + {}^3\mathrm{H} \rightarrow {}^3\mathrm{He} + e^-$, a reaction that has no energy threshold and can therefore occur even for the extremely low-energy neutrinos that make up the CNB. The signature of this process is a monoenergetic electron emitted with an energy above the endpoint of ordinary tritium beta decay, requiring sub-eV energy resolution and exceptionally low backgrounds. By combining a large tritium target with state-of-the-art spectrometry and calorimetry, PTOLEMY is designed to achieve the sensitivity needed to detect the expected handful of relic neutrino capture events per year. The predictions presented in this article for the axion decays into neutrinos could in principle be tested by these type of experiments, but since the energy of the neutrinos is larger, the detection rate can be much larger than in the CNB case. Therefore, one could be optimistic about the testability of these results even if it is very challenging.
%%%%%%%%%%%%%%%%%%%%%%%%
\section{SUMMARY}
%%%%%%%%%%%%%%%%%%%%%%%%
\label{sec:summary}

In this work, we presented a unified study of theoretical frameworks in which the solution to the strong CP problem, the origin of neutrino masses, and the nature of dark matter are all connected through a common symmetry-breaking scale. The central idea is that the spontaneous breaking of the PQ symmetry not only gives rise to the QCD axion, but also generates the heavy mass scales required for neutrino mass generation.
We analyzed several representative neutrino-mass mechanisms in both DFSZ and KSVZ realizations of the PQ symmetry, including the Type-I, Type-II, and Type-III seesaws, the Zee radiative model, and the colored seesaw. For each scenario, we determined the anomaly coefficients $E$ and $N$, derived the resulting axion couplings to photons and neutrinos, and studied the associated phenomenological consequences.

In the DFSZ framework, the Type-I seesaw, Type-II seesaw, and Zee model leave the electromagnetic and color anomaly structure unchanged relative to the minimal DFSZ construction, and therefore reproduce the standard DFSZ prediction for the axion-photon coupling. By contrast, the Type-III seesaw introduces electroweak fermion triplets that modify the electromagnetic anomaly coefficient and shift the ratio $E/N$, leading to a distinct and potentially testable prediction for $g_{a\gamma\gamma}$.

In the KSVZ framework, the axion-photon coupling is instead controlled by the gauge quantum numbers of the heavy vector-like colored fermions. 
We show for the first time the predictions in all possible scenarios with colored seesaw and show that in these scenarios using only the fields needed to generate neutrino masses at one-loop level, one can implement the Peccei-Quinn mechanism for the QCD axion. Notice that in other scenarios one can add extra fields to achieve the main goal but the answer is not unique as in the case of the colored seesaw scenarios.
 These models predict clearly separated values of $E/N$, and therefore distinct regions in the $g_{a\gamma\gamma}$--$m_a$ plane, providing promising targets for future haloscope searches and precision axion measurements.

We find that a generic prediction of all scenarios in which PQ breaking is tied to neutrino-mass generation is the existence of axion-neutrino couplings. These couplings are especially important for sufficiently heavy axions, where decays into neutrino pairs can compete with or dominate over the diphoton channel. As a result, they strongly affect the axion lifetime and branching fractions. In the low-mass regime, the diphoton decay remains dominant and the axion lifetime is far longer than the age of the Universe, preserving the viability of the QCD axion as a dark matter candidate.
If the QCD axion constitutes dark matter, the axion-neutrino interaction also implies a monochromatic neutrino flux from axion decays, peaked at $E_\nu = m_a/2$. We showed that this signal can lie above both the Cosmic Neutrino Background and the solar thermal neutrino background over a wide energy range, making it a potentially observable and complementary probe of the underlying PQ realization.

Taken together, our results show that linking PQ symmetry breaking to the origin of neutrino masses leads to a rich and correlated phenomenology across axion, neutrino, laboratory, astrophysical, and cosmological observables. The interplay among the axion-photon coupling, axion-neutrino coupling, axion lifetime, decay branching fractions, and monochromatic neutrino flux provides multiple complementary avenues to test these ideas. Future measurements could therefore not only establish the QCD axion as the solution to the strong CP problem, but also help identify the mechanism responsible for neutrino mass generation.

{\small{\textit{Acknowledgements}: We thank C. Murgui for discussions and comments on the manuscript. We thank the referee for very useful recommendations.}}
%%%%%%%%%%%%%%%%%%%%%%%%%%%%%%%%%%%%%%%%%%%%%%%%%%%
\begin{widetext}
\appendix
%%%%%%%%%%%%%%%%%%%%%%%%%%%%%%%%%%%%%%%%%%%%%%%%%%%
\section{Anomaly Coefficients for Colored Seesaw Mechanism}
\label{A}
The relevant Lagrangian for the colored seesaw mechanism is given by
\begin{eqnarray}
- \mathcal{L}
&\supset&
Y^\nu
\ell_L^T C i\sigma_2
\rho_L \Phi
+
\lambda_\rho
{\rm Tr}
(\rho^T_L C\rho_L)S 
+
\frac{\lambda}{2}
{\rm Tr}
(\Phi^\dagger H)^2
+\mathrm{h.c.},
\end{eqnarray}
where $\rho_L \sim (8, R, 0)$ with $R=1$ or 3. The $U(1)_{PQ}$ charge assignments are as follows,
\begin{equation}
    S\rightarrow S e^{i\chi_S}, \rho_{L} \rightarrow \rho_L e^{-i\frac{\chi_S}{2}}, \ell_L \rightarrow \ell_L e^{i\frac{\chi_S}{2}}, e_R \rightarrow e_R e^{i\frac{\chi_S}{2}}.
\end{equation}
 The $e_R$ charge is a consequence of the usual coupling term, $ Y_l\overline{\ell}_L H e_R$. We adopt the normalization $\chi_S=1$ without loss of generality because $U(1)_{PQ}$ is an Abelian symmetry without defined normalization.
 When the scalar field gets a VEV, the corresponding Lagrangian term becomes:
\begin{equation}
    \mathcal{L} \supset \frac{\lambda_{\rho} v_S}{\sqrt{2}} \ {\rm Tr}(\rho_L^T C \rho_L) e^{i\frac{a}{v_S}}.
\end{equation}
Now, working in the basis
\begin{equation}
    \rho_L \rightarrow \rho_L e^{-i\frac{a}{2v_S}},
\end{equation}
one can find the mass term for the $\rho_L$ field and the interactions of the axion to gluons and photons.
Using the Fujikawa's method one finds that the interaction term with gluons reads as
\begin{equation}
\mathcal{L}
    \supset \frac{g_s^2}{32 \pi^2}\frac{a}{v_S}  N \ G \tilde{G},
\end{equation}
where $N=d_C(F) T_D(F)$, $T_D(F)$ is the Dynkin index of the color fermionic representation (for $SU(3)_C$ representations, $T_D(3)= 1/2$ and $T_D(8)=3$), and $d_C (F)$ is the dimension of the fermionic representation.
In a similar way one finds the axion coupling to photons:
\begin{equation}
 \mathcal{L}
    \supset \frac{e^2}{32 \pi^2}\frac{a}{v_S} E  \ F \tilde{F},    
\end{equation}
where $E=\sum_i q_i^2 d_\gamma(F_i)$. Here $d_\gamma(F_i)$ is the dimension of the $F_i$ representation when one think about the contribution to the coupling to photons or the total contribution to this interaction of a given fermion representation. Now, we can show the predictions for the colored seesaw mechanisms. Here we have two main scenarios:

a) Colored Seesaw I: In this case we have $\rho_L \sim(8, 1, 0)$ with only neutral electrical fields, then we find
\begin{equation}
    E =0,
\ {\textrm{and}} \
    N = 3\times1=3.
\end{equation}

b) Colored Seesaw II: In this case we have $\rho_L \sim (8, 3, 0)$, which contains two electrically charged fields because it is a $SU(2)_L$ triplet with zero hypercharge, then one finds:
\begin{equation}
    E = [(-1)^2 +(1)^2]\times8=16, 
    \ {\textrm{and}} \
    N = 3\times3=9.
\end{equation}
%%%%%%%%%%%%%%%%%%%%%%%%%%%%%%%%%%
\section{Couplings to Neutrinos}
%%%%%%%%%%%%%%%%%%%%%%%%%%%%%%%%%%%
\subsection{Axion Neutrino Coupling in DFSZ Mechanism}
%%%%%%%%%%%%%%%%%%%%%%%%%%%%%%%%%%%%%%%%%%%%%%%%%%%%%%%%
Here we discuss the axion coupling to neutrinos in DFSZ scenarios.
For simplicity, let us discuss the simplest case with Type I seesaw. In this case, the relevant Lagrangian is given by
\begin{eqnarray}
\mathcal{L}_{\rm DFSZ}
&\supset&
Y_u \,\overline{q}_L H_u u_R
+
Y_d \,\overline{q}_L H_d d_R 
+
\lambda \,
H_u^T i\sigma_2 H_d\, {S^\dagger}^2
+
Y_\nu\,\overline{\ell}_L H_u \nu_R
+
\lambda_R \nu_R^T C \nu_R S
+\mathrm{h.c.}. 
\end{eqnarray}
Under the  $U(1)_{\mathrm{PQ}}$ charge assignments, the scalar fields have the following charge assignment:
\begin{equation}
    H_u \to e^{i X_{H_u}} H_u , \ H_d \to e^{i X_{H_d}}H_d \ \text{and} \ S \to e^{i X_S} S.
\end{equation}
After spontaneous symmetry breaking, the scalar fields can be written as 
\begin{equation}
    H_u \supset \frac{v_u}{\sqrt{2}} \begin{pmatrix}
        0 \\
        1
    \end{pmatrix} e^{i a_u/v_u}, \  H_d \supset \frac{v_d}{\sqrt{2}} \begin{pmatrix}
        0 \\
        1
    \end{pmatrix} e^{i a_d/v_d}\  \text{and} \  S \supset \frac{v_S}{\sqrt{2}}  e^{i a_s/v_S}.
\end{equation}
The axion field, $a$ can be written in terms of the pseudoscalar fields $a_u,a_d$ and $a_s$ as
\begin{equation}
    a= \frac{1}{v_a} \sum_i X_i v_i a_i  \hspace{0.3 cm}  \text{and} \ v_a^2 = \sum_i X_i^2 v_i^2. 
\end{equation}
 The charges are determined up to an overall normalization, which can be fixed by choosing a conventional value for $X_S$. In our convention 
\begin{equation}
    X_S =1, \hspace{1 cm} X_{H_u} = 2 (v_d/v)^2 , \hspace{1 cm} X_{H_d} = 2  (v_u/v)^2,
\end{equation}
where $\tan \beta = v_u/v_d$ and $v= \sqrt{v_u^2+v_d^2} \approx 246\  \text{GeV}$. Notice that, as  $v_S >> v_u ,v_d $, we have $v_a \approx v_S$.
When the Higgses ($H_u , H_d, S $) get VEV, the Lagrangian for the  DFSZ Type I becomes
\begin{align}
    -\mathcal{L}_{\text{DFSZ}}  \supset &  \  m_u \bar{u}_L u_R e^{i X_{H_u} a/v_a}\  + \  m_d  \bar{d}_L d_R e^{i X_{H_d} a/v_a}   
   + m_{\nu}^D \bar{\nu}_L \nu_R\ e^{i X_{H_u} a/v_a} + m_{R}  \nu_R^T C \nu_R  e^{i X_{S} a/v_a}  + {\rm{h.c.}}.
\end{align}
One can work on the mass basis redefining the fermion fields as 
\begin{align}
   & u_L \to e^{ i X_{H_u} a/(2 v_a)}\ u_L , \
    u_R \to e^{ -i X_{H_u} a/(2 v_a)}\ u_R, \\
   & d_L \to e^{i  X_{H_d}a/(2v_a)}\ d_L, \
    d_R \to e^{-i  X_{H_d}a/(2v_a)}\ d_R, \\
   & \nu_R \to e^{-i X_S a/(2v_a)}\  \nu_R , \
    \nu_L \to e^{i (2X_{H_u}-X_S) a/(2v_a)} \ \nu_L.
\end{align}
 In the DFSZ Type I scenarios, after the field redefinition of $\nu_L$, and from the kinetic term one finds
\begin{align}
i \bar{\nu}_L \gamma^{\mu} \partial_\mu \nu_L \to    
  (2X_{H_u}-X_S) \frac{\partial_\mu a}{4 v_a}  \bar{\nu} \gamma^\mu  \gamma^5\nu 
& = c_\nu \frac{\partial_\mu a}{4 v_a}  \bar{\nu} \gamma^\mu \gamma^5\nu.
\end{align}
Therefore,
$$c_\nu = (2 X_{H_u}-X_s)=-1 + 4/(1+\tan^2 \beta).$$
In the case when the neutrinos are on-shell, one can use the Dirac equation to show that 
\begin{equation}
    \partial_\mu (\bar{\nu} \gamma^\mu \gamma^5 \nu) = i2m_\nu \bar{\nu} \gamma^5 \nu,
\end{equation}
and find the coupling of the axion to one-shell neutrinos. The relevant interaction can then be written as 
\begin{eqnarray}
    \mathcal{L} \supset \frac{i}{2}{g_{a\nu\nu}} \ a \bar{\nu} \gamma^5\nu,
\end{eqnarray}
where
\begin{equation}
    v_a = N f_a \hspace{0.5 cm} \text{and} \hspace{0.5cm}  g_{a\nu_i \nu_i} = c_\nu \frac{m_{\nu_i}}{N f_a}.
\end{equation}
%%%%%%%%%%%%%%%%%%%%%%%%%%%%%%%%%%%%%%%%%%%%%%%%%%%%%%%
\subsection{Axion Neutrino Couplings in Colored Seesaw}
%%%%%%%%%%%%%%%%%%%%%%%%%%%%%%%%%%%%%%%%%%%%%%%%%%%%%%
In the KSVZ and colored seesaw scenarios, the relevant Lagrangian is given by
\begin{equation}
- \mathcal{L}
\supset
Y^\nu
\ell_L^T C i\sigma_2
\rho_L \Phi
+
\lambda_\rho
{\rm Tr}
(\rho^T_L C\rho_L)S + 
\frac{\lambda}{2}
{\rm Tr}
(\Phi^\dagger H)^2
+\mathrm{h.c.}.
\end{equation}
The scalar field gets a VEV it can be written as $S \supset \frac{v_S}{2}e^{ia_S/v_S}$. As a result, it is convenient to work in the mass basis where
\begin{align}
    &\rho_L \rightarrow \rho_L e^{-i\frac{a}{2v_S}}, \
    \ell_L \rightarrow \ell_L e^{i\frac{a}{2v_S}}, \
    e_R \rightarrow e_R e^{i \frac{a}{2v_S}}.
\end{align}
Now, from the kinetic term one finds
\begin{equation}
i \overline{\nu}_L \gamma^{\mu} \partial_{\mu} \nu_L \rightarrow i \overline{\nu}_L \gamma^{\mu} \partial_{\mu} \nu_L  -  \frac{\partial_{\mu} a}{4v_S} \overline{\nu} \gamma^{\mu}  \gamma^5 \nu.
\end{equation}
Therefore, $c_\nu =1$ in the colored seesaw scenarios.

\end{widetext}

\bibliography{refs}

\end{document}